\documentclass{emulateapj}
\usepackage{graphicx}
\usepackage{amssymb}
\usepackage{longtable}
\usepackage{epsf}
\usepackage{apjfonts}

\slugcomment{Accepted to the Astrophysical Journal}

\shorttitle{Rest-frame optical LF from $z=4$}
\shortauthors{Marchesini et al.}

\begin{document}

\title{The Evolution of the Rest-frame $V$-band Luminosity Function 
from $z=4$: \\ A Constant Faint-end Slope over the Last 12~Gyr of 
Cosmic History}

\author{Danilo Marchesini\altaffilmark{1},
  Mauro Stefanon\altaffilmark{2}, 
  Gabriel B. Brammer\altaffilmark{3}, and
  Katherine E. Whitaker\altaffilmark{4}}

\altaffiltext{1}{Department of Physics and Astronomy, Tufts University, 
Medford, MA 02155, USA}
\altaffiltext{2}{Observatori Astron\`omic Universitat de Val\`encia, C/ Catedr\'atico Agust\'in Escardino Benlloch, 7, 46980, Valencia, Spain} 
\altaffiltext{3}{European Southern Observatory (ESO), Santiago, Chile} 
\altaffiltext{4}{Department of Astronomy, Yale University, New Haven, CT 06520, USA} 

\begin{abstract}
We present the rest-frame $V$-band luminosity function (LF) of 
galaxies at $0.4 \leq z < 4.0$, measured from a near-infrared
selected sample constructed from the NMBS, the FIRES, the FIREWORKS, 
and the ultra-deep NICMOS and WFC3 observations in the HDFN, HUDF, 
and GOODS-CDFS, all having high-quality optical to mid-infrared data. 
This unique sample combines data from surveys with a large range of 
depths and areas in a self-consistent way, allowing us to (1) minimize 
the uncertainties due to cosmic variance; and (2) simultaneously constrain 
the bright and faint ends with unprecedented accuracy over the targeted 
redshift range, probing the LF down to $0.1~L^{\star}$ at $z\sim3.9$. We 
find that (1) the faint end is fairly flat and with a constant slope from 
$z=4$, with $\alpha=-1.27\pm0.05$; (2) the characteristic magnitude has 
dimmed by 1.3~mag from $z\sim3.7$ to $z=0.1$; (3) the characteristic 
density has increased by a factor of $\sim8$ from $z\sim3.7$ to $z=0.1$, 
with 50\% of this increase from $z\sim4$ to $z\sim1.8$; and (4) the 
luminosity density peaks at $z\approx1-1.5$, increasing by a factor of 
$\sim4$ from $z=4.0$ to $z\approx1-1.5$, and subsequently decreasing by 
a factor of $\sim1.5$ by $z=0.1$. We find no evidence for a steepening of 
the faint-end slope with redshift out to $z=4$, in contrast with previous 
observational claims and theoretical predictions. The constant faint-end 
slope suggests that the efficiency of stellar feedback may evolve with 
redshift. Alternative interpretations are discussed, such as different 
masses of the halos hosting faint galaxies at low and high redshifts and/or 
environmental effects.
\end{abstract}



\section{Introduction}

In the current paradigm of structure formation, dark matter (DM) halos
build up in a hierarchical fashion through the dissipationless
mechanism of gravitational instability. The assembly of the stellar
content of galaxies is instead governed by much more complicated
physical processes, often dissipative and non-linear, which are
generally poorly understood. To counter this lack of understanding, 
simplified analytical prescriptions of complex physical processes 
are employed in galaxy formation models. One of the fundamental tools 
to constrain the free parameters of these models is the luminosity 
function (LF), as its shape retains the imprint of galaxy formation 
and evolution processes.

The underlying physical processes that shape the faint end of the LF are 
generally associated with feedback from supernovae that is effective in 
heating gas and driving winds in shallow gravitational potentials 
\citep{dekel86}. Suppression of gas cooling in low-mass halos due to a 
background of photo-ionizing radiation also contributes in shaping the 
faint end (e.g., \citealt{benson02}). Matching the bright end of the LF has 
proven more challenging. Recent implementation of active galactic nucleus 
(AGN) feedback in semi-analytic models has yielded faithful reproductions 
of the observed local rest-frame optical and near-infrared global LFs 
(e.g., \citealt{granato04}; \citealt{bower06}; \citealt{croton06}; 
\citealt{menci06}; \citealt{kang06}; \citealt{monaco07}; 
\citealt{somerville08}). These models provide good matches to the LFs out 
to $z\sim1.5$ (e.g., \citealt{somerville11}; and references therein), 
although significant disagreements with observations are still present at 
$z \gtrsim 2$ (e.g., \citealt{henriques11}).

Within the hierarchical structure formation paradigm, one naturally expects 
strong evolution with redshift of the faint-end slope with steeper slopes 
at earlier time, considering that the slope of the dark matter mass function 
is very steep ($\alpha_{\rm DM}\sim -2$), and that the objects that form in 
these halos continue to grow by continued star formation and mergers with 
each other (e.g., \citealt{khochfar01}), hence flattening the slope.

Observationally, evidence of evolution of the faint-end slope $\alpha$ has 
been recently claimed, with $\alpha$ steepening with redshift \citep{ryan07}. 
\citet{khochfar07} have shown that simulations of galaxy formation and 
evolution are able to reproduce the observed redshift evolution of $\alpha$. 
Specifically, while supernova feedback is responsible for flatter $\alpha$ 
with respect to $\alpha_{\rm DM}$ at any given redshift, its evolution with 
redshift appears to be driven by a steeper dark matter mass function at 
early cosmic times for the range of halo masses hosting sub-$L^{\star}$ 
galaxies  (e.g., \citealt{khochfar07}; \citealt{kobayashi07}).

Despite the significant improvements in constraining the faint-end of the 
rest-frame UV LF of dropout galaxies out to $z\sim8$ (e.g., \citealt{oesch10b};
\citealt{bouwens11}; and references therein), the faint-end slope of the 
rest-frame optical LFs of galaxies at $z\gtrsim1$ from near-infrared (NIR) 
selected samples, which arguably allow for the assembly of representative 
samples of high-redshift galaxies, is still quite uncertain. Moreover, 
the evidence of a steepening of $\alpha$ with redshift comes from several 
surveys, each with unique selection effects, observational biases, and 
measured at different rest-frame wavelengths. Specifically, all measurements 
at $z>3.5$ come from studies in the rest-frame far-UV, whereas at $z\lesssim3$ 
LFs are typically measured in the rest-frame optical.

In this paper, we derive the rest-frame optical ($V$-band) LFs of galaxies 
over the redshift interval $0.4 \leq z<4.0$ using a NIR-selected composite 
sample of galaxies constructed from the NMBS, the FIRES, the GOODS-CDFS 
surveys, and ultra-deep {\it Hubble Space Telescope} (HST) data that 
self-consistently combines the advantages of deep, pencil beam surveys with 
those of shallow, wider surveys. This NIR composite sample allows us to 
(1) minimize the uncertainties due to cosmic variance, and empirically 
quantify its contribution to the total error budget by exploiting the large 
number of independent field of views; and (2) simultaneously constrain the 
bright and faint ends of the LF with unprecedented accuracy and statistics  
over the entire targeted redshift range. This paper is structured as follows. 
In Section~\ref{sample}, we present the composite NIR-selected sample used to 
measure the LFs of galaxies at $0.4 \leq z < 4.0$; the methods used to measure 
the LF (the $1/V_{\rm max}$ and the maximum likelihood methods) are presented 
in Section~\ref{lf}, as well as the LFs of galaxies at $0.4\leq z<0.7$, 
$0.76\leq z<1.1$, $1.1\leq z<1.5$, $1.5\leq z<2.1$, $2.1\leq z<2.7$, 
$2.7\leq z<3.3$, and $3.3\leq z<4.0$, and the evolution of the Schechter 
function parameters and luminosity density. Our results are summarized in 
Section~\ref{conc}. We assume $\Omega_{\rm M}=0.3$, $\Omega_{\rm \Lambda}=0.7$, 
and $H_{\rm 0}=70$~km~s$^{-1}$Mpc$^{-1}$. All magnitudes are in the AB system.


\section{The NIR Composite Sample} \label{sample}

The data set we use to measure the LF consists of a composite 
NIR-selected sample of galaxies built from several multi-wavelength 
surveys, all having high-quality UV to mid-infrared photometry: 
the Faint InfraRed Extragalactic Survey (FIRES; 
\citealt{franx03}), the Great Observatories Origins Deep Survey 
(GOODS; \citealt{giavalisco04}), the NEWFIRM Medium-Band Survey 
(NMBS; \citealt{vandokkum09}), the ultra-deep NICMOS observations 
over the HDF-North (HDFN) GOODS field (\citealt{dickinson99}; 
\citealt{thompson99}), the ultra-deep WFC3/IR observations 
taken as part of the HUDF09 program (GO~11563; PI: Illingworth) over 
the Hubble Ultra-Deep Field (HUDF), and the wide-area WFC3 Early 
Release Science (ERS) observations over the CDF-South GOODS field 
(GO~11359; PI: O'Connell). 

For the FIRES, GOODS-CDFS, and NMBS surveys we have adopted the 
publicly-available $K$-band selected catalogs from \citet{labbe03}, 
\citet{wuyts08}, and \citet{whitaker11}, respectively. We refer to these 
works for a complete description of the observations, reduction procedures, 
and the construction of the $K$-selected multiwavelength catalogs. Briefly, 
the FIRES-HDFS catalog (HDFS, hereafter) has 833 sources down to 
$K_{\rm S}^{\rm tot}=26.0$ over an area of 2.5$^{\prime} \times$2.5$^{\prime}$. 
The FIREWORKS-CDFS (CDFS, hereafter) K$_{\rm S}$-selected catalog comprises 
6308 objects down to $K_{\rm S}^{\rm tot}=24.6$ over a total surveyed area of 
138~arcmin$^{2}$. The NMBS (COSMOS and AEGIS, hereafter) K-selected catalogs 
comprise 31,306 and 27,572 objects, respectively, down to 
$K_{\rm S}^{\rm tot}=24.5$ and over a total surveyed area of 0.5~deg$^{2}$. 

For the remaining fields, we have constructed $H_{\rm 160}$-selected catalogs 
following previous studies (e.g., \citealt{labbe03}). The full details of 
the reduction, source detection, and generation of the photometric catalogs 
will be described in Stefanon et al. (2012, in prep.). Briefly, similarly to 
the other adopted catalogs, we have used SExtractor \citep{bertin96} in 
dual-image mode, with fluxes measured in the registered and PSF-matched 
images in order to derive accurate colors and limit any bandpass-dependent 
effects. We have used an alternative source fitting algorithm especially 
suited for heavily confused images for which a higher resolution prior (in 
this case, the $H_{\rm 160}$-band image) is available to extract the photometry 
from the IRAC images. This method is described in more detail in the appendix 
of \citet{marchesini09}. The UDF WFC3 (UDF, hereafter), GOODS-CDFS WFC3 (ERS, 
hereafter), and the HDFN NICMOS (HDFN, hereafter) H$_{\rm 160}$-selected 
catalogs comprise 2100, 9470, and 1774 objects down to $H_{\rm 160}=28.5$, 
$H_{\rm 160}=27.3$, and $H_{\rm 160}=28.0$, over an area of 5.3, 46.2, and 6.7 
arcmin$^{2}$, respectively. 

Table~\ref{tab-sample} provides a summary of the catalogs used, including 
the filters that each field has been imaged with, the observed $K$- and 
$H$-band limiting magnitudes adopted to construct the composite sample 
($K^{\rm tot}_{\rm lim}$ and $H^{\rm tot}_{\rm lim}$; see \S~\ref{completeness}), 
the effective area of each pointing ($A_{\rm eff}$), the number of galaxies 
over the redshift interval $0.4<z<4.0$ used to measure the rest-frame $V$-band 
LFs ($N_{\rm 0.4<z<4.0}$), and the corresponding fraction of galaxies with 
spectroscopic redshift ($f_{\rm zspec}$). For the HST $H_{\rm 160}$-selected 
catalogs, $A_{\rm eff}$ represents the area in the fields where the NIR 
observations are at least 10\% of their maximum depths in the field, and 
with full coverage in the ACS bands. For the HDFS and FIREWORKS catalogs, 
$A_{\rm eff}$ represents the area in the fields over which the optical and NIR 
observations are at least 20\% of their maximum depths in the field, following 
suggested criteria by \citet{labbe03} and \citet{wuyts08}.\footnote{The area 
of FIREWORKS listed in Table~\ref{tab-sample} excludes the area covered by 
the ERS and the UDF catalogs.} Finally, for the 
NMBS catalogs, $A_{\rm eff}$ represents the area in the fields over which the 
optical and NIR observations are at least 30\% of their maximum depths in the 
field, again following suggested criteria by \citet{whitaker11}. The regions 
around bright stars were also excluded in the calculation of the LF, as faint 
galaxies in these regions are either obscured by the foreground star or have 
systematically incorrect magnitudes due to scattered light.

\subsection{Completeness} \label{completeness}

The $H_{\rm 160}$- and $K$-band limiting magnitudes listed in 
Table~\ref{tab-sample} were chosen following a conservative approach that 
ensures completeness better than 90\% in all catalogs down to the adopted 
limiting magnitudes. The completenesses for the FIRES and FIREWORKS surveys 
have been characterized in \citet{labbe03} and \citet{wuyts08}. For these 
catalogs, we have used the same $K$-band limiting magnitudes previously 
adopted in \citet{marchesini09}.

The completenesses of the NMBS catalogs have been derived in 
\citet{whitaker11}. When masking sources, the point-source completeness is 
better than 90\% at $K^{\rm tot}_{\rm S}<23.2$. When objects are not masked, 
objects can fall on or close to other sources, and these blended objects are 
not properly handled by SExtractor. In this case, the completeness in COSMOS 
and AEGIS is better than 85 and 87\% at $K^{\rm tot}_{\rm S}<22.7$ and 22.8, 
respectively. However, in our work we have adopted the ``de-blended'' catalogs, 
for which these completeness limits are quite conservative \citep{whitaker11}. 
To derive robust limiting magnitudes for the NMBS de-blended catalogs used in 
our work, we have used the raw number counts of galaxies, which do not show 
any signs of incompleteness down to $K^{\rm tot}_{\rm S} \approx 22.8$, implying 
completeness better than 95\% in the de-blended NMBS catalogs at 
$K^{\rm tot}_{\rm S}<22.7$ and 22.8 for COSMOS and AEGIS, respectively. 
Therefore, we have adopted $K^{\rm tot}_{\rm S}=22.7$ and 
$K^{\rm tot}_{\rm S}=22.8$ for COSMOS and AEGIS, respectively, as the limiting 
magnitudes to construct the NIR-selected composite sample. 

The completenesses for the HST $H_{\rm 160}$-selected catalogs (UDF, HDFN, and 
ERS) will be presented in detail in Stefanon et al. (2012; in prep.). Briefly, 
the completeness in each catalog was derived following the method in 
\citet{whitaker11}. The (point-source) completeness was estimated as a 
function of magnitude by attempting to recover simulated sources within the 
$H_{\rm 160}$-band noise-equalized detection image. The derived completeness 
is better than 90\% at $H_{\rm 160}<$28.2, 26.7, and 26.6 for the UDF, ERS, 
and HDFN catalogs, respectively. We have therefore adopted the conservative 
limiting magnitudes listed in Table~\ref{tab-sample}, to construct the 
NIR-selected composite sample.

\begin{deluxetable*}{lclccccc}
\centering
\tablecaption{Summary of catalogs used to construct the NIR-selected composite sample \label{tab-sample}}
\tablehead{
  \colhead{Field/Survey} & \colhead{Selection Band} & 
  \colhead{Filter Coverage} & \colhead{$K/H^{\rm tot}_{\rm lim}$} &
  \colhead{$A_{\rm eff}$} & \colhead{$N_{\rm 0.4<z<4.0}$} & \colhead{$f_{\rm zspec}$} & 
  \colhead{Ref.} \\
  \colhead{}             &  \colhead{}              & 
  \colhead{}                & (mag)                         &
  (arcmin$^{2}$)        &     \colhead{}           &      [\%]          & 
  \colhead{} }
\startdata
UDF        & WFC3 H$_{\rm 160}$   & ACS~B$_{\rm 435}$V$_{\rm 606}$i$_{\rm 775}$z$_{\rm 850}$ & 27.8 & 5.1 & 547 & 7.9 & (1,2,3,4)\\
           &                        & NICMOS~J$_{\rm 110}$H$_{\rm 160}$                  &       &     &     &     &        \\
               &                        & WFC3~Y$_{\rm 105}$J$_{\rm 125}$H$_{\rm 160}$          &       &     &     &     &        \\
              &                        & ISAAC~K$_{\rm S}$, IRAC                 &      &     &     &     &        \\
ERS         & WFC3 H$_{\rm 160}$   & ACS~B$_{\rm 435}$V$_{\rm 606}$i$_{\rm 775}$z$_{\rm 850}$   & 26.3 & 42.5 & 2437 & 10.1 & (3,4,5,6,7)\\
            &                        & WFC3~UV$_{\rm 225, 275, 336}$Y$_{\rm 105}$J$_{\rm 125}$H$_{\rm 160}$ & &     &     &     &  \\
             &                        & ISAAC~HK$_{\rm S}$, IRAC                  &      &     &     &     &        \\
HDFN       & NICMOS H$_{\rm 160}$ & ACS~B$_{\rm 435}$V$_{\rm 606}$i$_{\rm 775}$z$_{\rm 850}$ & 26.1 & 6.7 & 435 & 15.6 & (4,5,6,8,9,10)\\
              &                        & WFPC2~U$_{\rm 300}$B$_{\rm 450}$V$_{\rm 606}$I$_{\rm 814}$ &      &     &     &     &        \\
             &                        & NICMOS~J$_{\rm 110}$H$_{\rm 160}$, IRAC              &      &     &     &     &        \\
HDFS/FIRES     & ISAAC K$_{\rm s}$     & WFPC2~U$_{\rm 300}$B$_{\rm 450}$V$_{\rm 606}$I$_{\rm 814}$ & 25.6 & 4.5 & 204 & 8.3 & (11,12,13)\\
              &                        & ISAAC~J$_{\rm S}$HK$_{\rm S}$, IRAC            &      &     &     &     &        \\
CDFS/FIREWORKS & ISAAC K$_{\rm s}$     &  ACS~B$_{\rm 435}$V$_{\rm 606}$i$_{\rm 775}$z$_{\rm 850}$ & 23.2 & 77.2 & 689 & 44.6 & (5,14)\\
              &                        & WFI~U$_{\rm 38}$BVRI, IRAC            &      &     &     &     &        \\
            &                        & ISAAC~J$_{\rm S}$HK$_{\rm S}$, IRAC~CH1:4            &      &     &     &     &        \\
AEGIS/NMBS      & NEWFIRM K         & GALEX~FUV$_{1500}$NUV$_{2300}$    & 22.8 & 743 & 7402 & 15.9 & (15,16)\\
                 &                        & CFHTLS~ugriz          &      &     &     &     &        \\
             &                        & NEWFIRM~J$_{\rm 1}$J$_{\rm 2}$J$_{\rm 3}$H$_{\rm 1}$H$_{\rm 2}$K          &      &     &     &     &        \\
              &                        & WIRCam~JHK$_{\rm S}$, IRAC          &      &     &     &     &        \\
COSMOS/NMBS      & NEWFIRM K         & GALEX~FUV$_{1500}$NUV$_{2300}$     & 22.7 & 741 & 7689 & 7.3 & (15,16)\\
            &                        & Subaru~B$_{\rm J}$V$_{\rm J}$r$^{\prime}$i$^{\prime}$z$^{\prime}$, CFHTLS~ugriz   &   &   &  &     &       \\
            &                        & Subaru~12 medium-bands         &      &     &     &     &        \\
             &                        & NEWFIRM~J$_{\rm 1}$J$_{\rm 2}$J$_{\rm 3}$H$_{\rm 1}$H$_{\rm 2}$K          &      &     &     &     &        \\
             &                        & WIRCam~JHK$_{\rm S}$, IRAC          &      &     &     &     &        
\enddata
\tablecomments{$K/H^{\rm tot}_{\rm lim}$ is the observed $K$- and 
$H_{\rm 160}$-band limiting magnitudes adopted to construct the 
NIR-selected composite sample; $A_{\rm eff}$ is the effective area; 
$N_{\rm 0.4<z<4.0}$ is the number of galaxies over the redshift interval 
$0.4 \leq z < 4.0$ used to measure the rest-frame $V$-band LFs, and 
$f_{\rm zspec}$ is the corresponding fraction of galaxies with $z_{\rm spec}$.
The references of the used catalogs and data sets are 
(1) \citet{thompson05}, (2) \citet{beckwith06}, 
(3) \citet{retzlaff10}, (4) Stefanon et 
al. (2012; in prep.), (5) \citet{giavalisco04}, 
(6) \citet{dickinson03}, (7) \citet{damen11}, 
(8) \citet{williams96}, (9) \citet{dickinson99}, 
(10) \citet{thompson99}, (11) \citet{franx03}, 
(12) \citet{labbe03}, (13) \citet{wuyts07}, 
(14) \citet{wuyts08}, (15) \citet{vandokkum09}, 
(16) \citet{whitaker11}.}
\end{deluxetable*}

\subsection{Photometric Redshifts} \label{photoz}

\begin{figure}
\epsscale{0.8}
\plotone{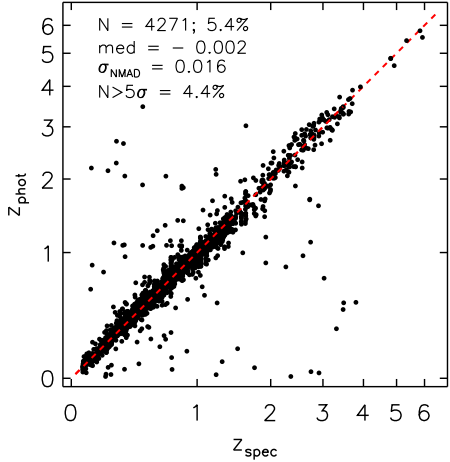}
\caption{Spectroscopic vs. photometric redshifts for the NIR-selected 
catalogs, shown on a ``pseudo-log'' scale. The number and fraction 
of sources with spectroscopic redshifts used in the shown comparison are 
also specified, as well as the median in $\Delta z / (1+z_{\rm spec})$, 
$\sigma_{\rm NMAD}$, and the fraction of catastrophic outliers. The comparison 
between $z_{\rm spec}$ and $z_{\rm phot}$ is extremely good, both at low and high 
redshifts. \label{fig1b}}
\end{figure}

The overall fraction of sources with spectroscopic redshifts in the 
$H_{\rm 160}$- and $K$-selected catalogs is 5.4\%. Consequently, we must rely 
primarily on photometric redshift estimates. Photometric redshifts 
$z_{\rm phot}$ for all galaxies 
were derived using the EAzY photometric redshift code \citep{brammer08}, 
adopting $z\_peak$ as the photometric redshift, which finds discrete peaks 
in the redshift probability function and returns the peak with the largest 
integrated probability. The template set used in this work is described in 
\citet{whitaker11}, composed of EAzY default templates and an additional 
template for an old, red galaxy. We quantify the accuracy of the photometric 
redshifts $z_{\rm phot}$ by comparing them to the available spectroscopic 
redshifts $z_{\rm spec}$, as shown in Figure~\ref{fig1b}.

The spectroscopic redshifts for the NMBS COSMOS and AEGIS catalogs were 
taken from the $z$COSMOS survey \citep{lilly07} and the Deep Extragalactic 
Evolutionary Probe 2 \citep{davis03}, respectively. Spectroscopic redshifts 
of 117 Lyman break galaxies at $z\sim 3$ within the AEGIS field were also 
added from \citet{steidel03}. For FIRES and FIREWORKS we used the spectroscopic 
redshifts compiled in \citet{rudnick03} and \citet{wuyts08}, respectively. 
For the UDF and ERS $H_{\rm 160}$-selected catalogs, spectroscopic redshifts 
were taken from the ``CDFS Master Catalog'' of spectroscopic redshifts (v2.0; 
I. Balestra; December 2009).\footnote{\url{http://www.eso.org/sci/activities/projects/goods/MasterSpectroscopy.html}} The spectroscopic redshifts for the HDFN 
$H_{\rm 160}$-selected catalog were taken from \citet{barger08}.

\begin{figure}
\epsscale{1.1}
\plotone{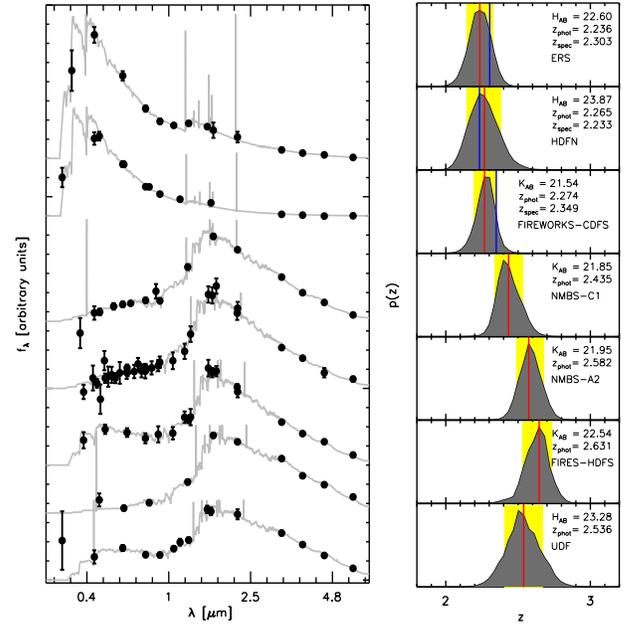}
\caption{{\it Left:} SEDs of galaxies at $2.1<z<2.7$ taken from 
the seven NIR-selected catalogs used to construct the composite 
sample: ERS, HDFN, FIREWORKS CDFS, NMBS COSMOS, NMBS AEGIS, FIRES 
HDFS, and UDF (from top to bottom). The galaxies were arbitrarily 
selected to have photometric redshifts falling in the targeted 
redshift interval, with reasonably well-behaved redshift probability 
functions. They are not representative of the galaxy population at 
$2.1<z<2.7$. Circles are the observed fluxes in units of $F_{\lambda}$, 
with 1~$\sigma$ errors. The solid gray curves represent the best-fit 
EAzY templates; each SED was normalized and offset with respect to the 
others.{\it Right:} EAzY redshift probability functions (gray regions). 
The red line is the adopted redshift from EAzY ($z_{\rm peak}$); the blue 
line is the spectroscopic redshift, when available. The yellow regions 
are the 1~$\sigma$ allowed values for $z_{\rm peak}$. Also listed are the 
$K$- or $H$-band total magnitudes, $z_{\rm peak}$, and $z_{\rm spec}$.
\label{fig1}}
\end{figure}

The comparison between spectroscopic and photometric redshifts results in 
excellent $z_{\rm phot}$ estimates. The median in $\Delta z / (1+z_{\rm spec})$, 
with $\Delta z = z_{\rm phot} - z_{\rm spec}$, is -0.000, -0.002, -0.007, 0.018, 
-0.005, -0.009, and -0.015 in COSMOS, AEGIS, CDFS, HDFS, HDFN, ERS, and UDF, 
respectively, with the normalized median absolute deviation 
$\sigma_{\rm NMAD}$\footnote{The normalized median absolute deviation 
$\sigma_{\rm NMAD}$, defined as 
$1.48 \times median[|(\Delta z - median(\Delta z))/(1+z_{\rm spec})|]$, is 
equal to the standard deviation for a Gaussian distribution, and it is less 
sensitive to outliers than the usual definition of the standard deviation 
(e.g. \citealt{brammer08})}=0.008, 0.015, 0.029, 0.040, 0.029, 0.037, 0.039. 
The fraction of catastrophic outliers (defined as galaxies with 
$\Delta z / (1+z_{\rm spec})>5 \sigma_{\rm NMAD}$) is 4.8\%, 3.2\%, 2.9\%, 
1.9\%, 2.4\%, 3.6\%, and 4.6\% in COSMOS, AEGIS, CDFS, HDFS, HDFN, ERS, and 
UDF, respectively. Figure~\ref{fig1} provides examples of spectral energy 
distributions (SEDs) of galaxies at $2.2<z<2.7$ in the seven NIR-selected 
catalogs, along with their corresponding EAzY redshift probability 
functions. 

The accuracy and reliability of the photometric redshifts estimated for the 
NMBS survey have been discussed in detail in \citet{whitaker11}, where 
excellent agreement with the available spectroscopic redshifts is found. The 
accuracy of the photometric redshifts estimated in the FIRES and FIREWORKS 
catalogs have been discussed in detail in \citet{brammer08}, \citet{wuyts08}, 
and \citet{marchesini09}. A detailed discussion of the quality of the 
photometric redshifts in the $H_{\rm 160}$-selected catalogs in UDF, ERS, and 
HDFN will be presented in Stefanon et al. (2012; in prep.).

It should be stressed that the assessment of the errors of the photometric 
redshifts from the $z_{\rm phot}$ versus $z_{\rm spec}$ comparison works only if 
the spectroscopic sample is a random subset of the photometric sample. This 
is not true in general, as the currently available spectroscopic samples are 
heavily weighted toward bright and/or blue, low-redshift galaxies. However, 
\citet{brammer08} showed explicitly that the errors in $z_{\rm phot}$ versus 
$z_{\rm spec}$ plots are generally consistent with the errors estimated from 
both Monte Carlo simulations and from the formal errors in the fit, i.e., the 
confidence intervals output by EAzY correctly describe the deviations from 
the spectroscopic redshifts. We are therefore confident that the photometric 
redshift uncertainties returned by EAzY are reliable.

\subsection{Rest-frame Luminosities} \label{restframelum}

Rest-frame luminosities are derived by integrating the redshifted 
rest-frame filter bandpass from the best-fit EAzY template, as described 
in \citet{brammer11}. This method produces very similar rest-frame 
luminosities as other methods interpolating between the observed bands 
that bracket the rest-frame band at a given redshift (e.g., 
\citealt{rudnick03}). Specifically, down to the $H_{\rm 160}$- and $K$-band
limiting magnitudes of the NIR-selected composite sample, the difference 
between the rest-frame $V$-band magnitudes derived with the two methods 
are very small, $\sim0.02$~mag, and no dependence on the $H_{\rm 160}$- or 
$K$-band magnitude. We note that most galaxies in the NIR-selected composite 
sample have robust photometry in the IRAC bands\footnote{For the NIR-selected 
composite sample, 4.1\% of the sources at $0.4 \leq z < 4.0$ have IRAC 
signal-to-noise ratio (S/N) $<$3 in the 3.6~$\mu m$ band. This fraction is 
only slightly larger (5.5\%) in the highest targeted redshift interval 
$3.3 \leq z < 4.0$.}, which allows us to properly model the rest-frame optical 
SEDs of galaxies and to robustly derive their rest-frame $V$-band magnitudes 
out to the highest targeted redshift. We computed rest-frame luminosities in 
the $V$ passband defined by \citet{maiz06}. In all cases where a spectroscopic 
redshift is available, we computed the rest-frame luminosity fixed at 
$z_{\rm spec}$. 

\subsection{The Composite Sample} \label{compsample}

We constructed a composite NIR-selected sample of galaxies over the 
redshift range $0.4 \leq z<4.0$ to be used in the derivation of the 
rest-frame $V$-band LFs of galaxies in \S~\ref{lf}. The final composite 
sample includes 19403 NIR-selected galaxies with $H_{\rm 160}^{\rm tot}<27.8$ 
or $K^{\rm tot}<25.6$, and $0.4 \leq z<4.0$ over an effective area of 
1620~arcmin$^{2}$. Of these sources, 12.5\% have spectroscopic redshifts. This 
NIR-selected sample is unique in that it combines data from surveys with a 
large range of depths and areas in a self-consistent way. The wide, fairly 
deep, NMBS survey, with its exquisite photometric redshifts, provides 
optimal sampling of the bright end of the LF down to the characteristic 
magnitude at $z\sim3$ and a factor of $\sim20$ fainter than the characteristic 
luminosity, $L^{\star}$, at $z\sim0.6$. The ultra-deep $H_{\rm 160}$-selected 
catalogs allowed us to probe the faint end down to 0.1~$L^{\star}$ at 
$z\sim3.9$, and to 0.003~$L^{\star}$ at $z\sim1$ (see also Figure~\ref{fig2}).


\section{The Luminosity Function and Density} \label{lf}

\subsection{Methodology and Uncertainties}

To estimate the rest-frame $V$-band LF in the case of a composite sample, 
we have used the extended version of the $1/V_{\rm max}$ estimator as defined 
in \citet{avni80}. The Poisson error in each magnitude bin was computed 
adopting the recipe of \citet{gehrels86}. The $1/V_{\rm max}$ estimator has 
the advantages of simplicity and no a priori assumption of a functional 
form for the LF; it also yields a fully normalized solution. However, it 
can be affected by the presence of clustering in the sample. Uncertainties 
due to field-to-field variations in the determination of the LF with the 
$1/V_{\rm max}$ method were estimated following the procedure presented in 
\citet{marchesini07}, and added in quadrature to the Poisson errors. The 
contribution from cosmic variance to the error budget is typically 0.06~dex 
at the characteristic magnitude in all targeted redshift intervals, except 
at $3.3 \leq z<4.0$, where it increases to 0.1~dex.

We also measured the LF using the STY method \citep{sandage79}, which is a 
parametric maximum likelihood estimator. The STY method is unbiased with 
respect to density inhomogeneities (e.g., \citealt{efstathiou88}), it has 
well-defined asymptotic error properties (e.g., \citealt{kendall61}), and 
it does not require binning of the data. For the STY method, we have assumed 
that the LF is described by a \citet{schechter76} function,

\begin{eqnarray}
\Phi (M) = (0.4 \ln{10}) \Phi^{\star} 
\big[ 10^{0.4 (M^{\star}-M)(1+\alpha)} \big] {} \nonumber\\
 \cdot \exp{\big[ -10^{0.4(M^{\star}-M)} \big]},
\end{eqnarray}
where $\alpha$ is the faint-end slope parameter, $M^{\star}$ is the 
characteristic absolute magnitude, and $\Phi^{\star}$ is the normalization. 
The best-fit solution is obtained by maximizing the likelihood $\Lambda$ 
with respect to the parameters $\alpha$ and $M^{\star}$. The value of 
$\Phi^{\star}$ is then obtained by imposing a normalization on the best-fit 
LF such that the total number of observed galaxies in the sample is 
reproduced. We have added in quadrature the contribution from cosmic 
variance previously derived (i.e., 0.06~dex at $z<3.3$ and 0.1~dex at 
$3.3 \leq z<4.0$) to the error budget of $\Phi^{\star}$.

The uncertainties on the LF due to photometric redshift errors have been 
estimated following the recipe in \citet{marchesini09, marchesini10} . 
Briefly, for each galaxy in the NIR-selected sample, a set of 200 mock SEDs 
was created by perturbing each flux point according to its formal error bar. 
Second, photometric redshifts and rest-frame $V$-band luminosities were 
estimated as described in \S~\ref{sample}. Finally, the LFs were derived with 
the $1/V_{\rm max}$ and the STY method for each of the 200 Monte Carlo 
realizations of the NIR-selected sample. This approach naturally addresses 
the fact that fainter sources tend to be characterized by less accurate 
$z_{\rm phot}$ estimates due to the larger errors in their photometry, as well 
as sources characterized by power-law SEDs and consequently by very poorly 
constrained $z_{\rm phot}$ estimates and very broad $z_{\rm phot}$ distributions 
derived from the Monte Carlo realizations.\footnote{We note that catastrophic 
outliers in the photometric redshift distribution could potentially cause 
systematic errors in the LF measurements (e.g., \citealt{chen03}). 
\citet{marchesini07} performed extensive simulations to assess the impact 
of catastrophic outliers on the derived Schechter function parameters, 
showing that catastrophic outliers will result in measured LFs that are 
typically steeper than the true LFs, and with brighter characteristic 
luminosities. For example, at $z>2$, they found $\Delta \alpha = -0.01$ and 
$\Delta M^{\star} = -0.05$~mag when the effect of 5\% catastrophic outliers is 
simulated, much smaller than the random uncertainties on the Schechter 
function parameters. Whereas it is likely that these systematics effects 
get larger at lower redshift, especially for the characteristic luminosity, 
by scaling the results from the simulations in \citet{marchesini07}, we expect 
these systematic effects to be in general smaller than, or at most comparable 
to, the derived random uncertainties on the Schechter function parameters at 
$z<1$.}

The rest-frame $V$-band LFs of galaxies were derived in the following 
redshift intervals: $0.4\leq z<0.7$, $0.76\leq z<1.1$, $1.1\leq z<1.5$, 
$1.5\leq z<2.1$, $2.1\leq z<2.7$, $2.7\leq z<3.3$, and $3.3\leq z<4.0$. 
For all but the lowest redshift interval, the full NIR-selected composite 
sample was used. In the redshift interval $0.4\leq z<0.7$, only the NMBS 
$K$-selected samples were used, as the volumes probed by the other surveys 
become too small to be meaningful. We have excluded the redshift interval 
$0.7 \leq z<0.76$, as known clusters of galaxies are present in the COSMOS 
and CDFS fields (e.g., \citealt{guzzo07}; \citealt{trevese07}; 
\citealt{salimbeni09}), contaminating the field galaxy LFs presented in this 
paper.

Because of the coupling between the two parameters $\alpha$ and $M^{\star}$, 
the luminosity density (obtained by integrating the LF over all magnitudes) 
is a robust way to characterize the evolution of the LF with redshift. The 
luminosity density $\rho_{\rm L}$ was calculated using 

\begin{equation}
\rho_{\rm L} = \int^{\infty}_{0}{L_{\rm \nu} \Phi(L_{\rm \nu}) dL_{\rm \nu}} = \Gamma(2+\alpha)\Phi^{\star}L^{\star},
\label{eq-rho}
\end{equation}
which assumes that the Schechter parameterization of the observed LF is a 
good approximation and valid also at luminosities fainter than probed by our 
composite sample. The uncertainties on the luminosity density were calculated 
using the approach adopted in \citet{marchesini09}. Specifically, the 
1~$\sigma$ errors on the total luminosity density have been estimated by 
deriving the distribution of all of the values of $\rho_{\rm L}$ allowed within 
the 1~$\sigma$ solutions of the Schechter function parameters from the 
maximum likelihood analysis. The contributions to the total error budget from 
photometric redshift random uncertainties (derived with the Monte Carlo 
simulations) and from cosmic variance (i.e., 0.06~dex at $z<3.3$ and 0.1~dex 
at $3.3 \leq z<4.0$) were also added in quadrature.

\subsection{Results} \label{reslf}

\begin{figure*}
\epsscale{1.1}
\plotone{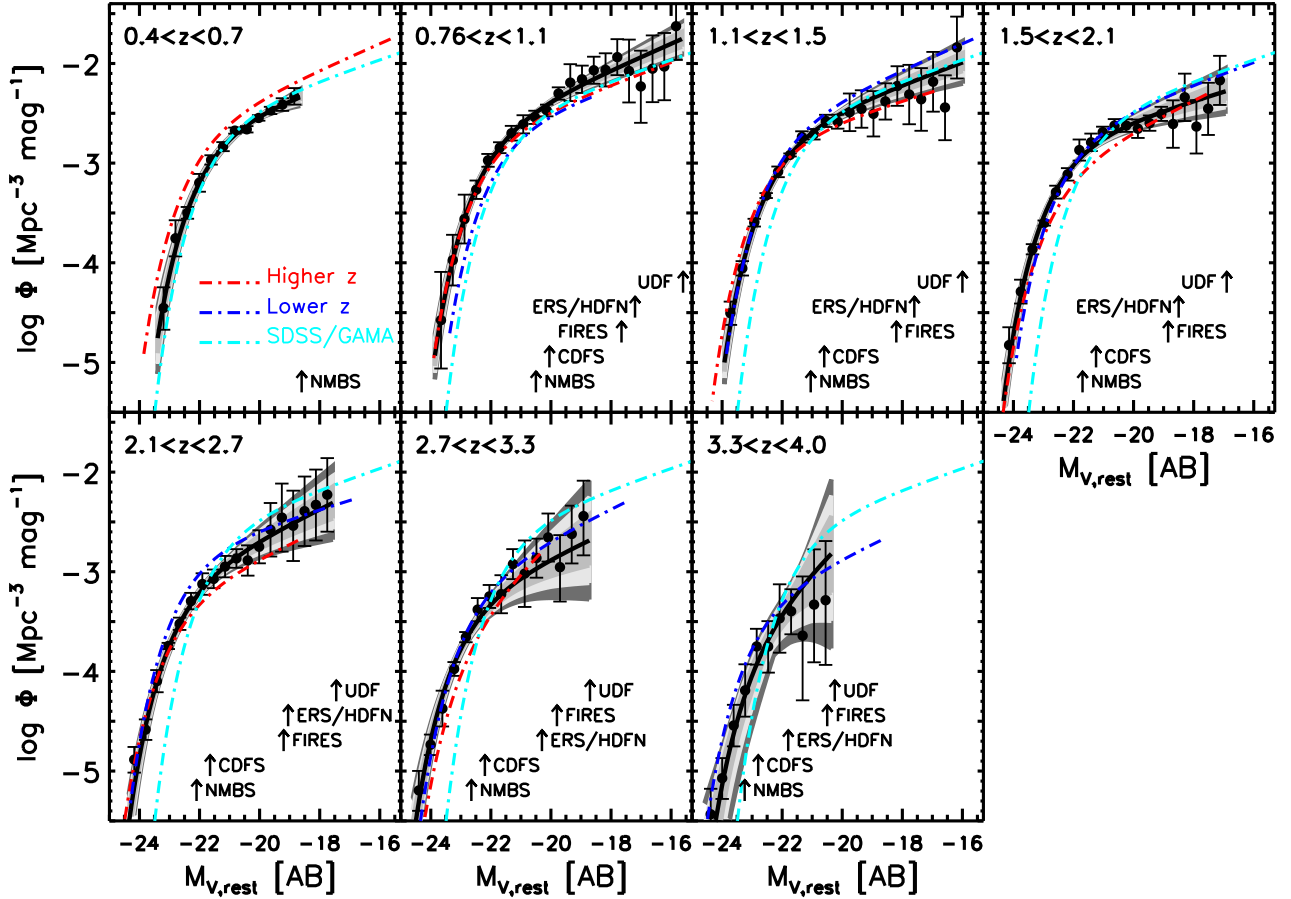}
\caption{Rest-frame $V$-band LFs of galaxies at $0.4\leq z < 4.0$. 
Black filled circles are the $1/V_{\rm max}$ method estimates with 
total 1-$\sigma$ error bars. The black solid curves are the LFs 
estimated with the maximum-likelihood method. The shaded gray 
regions represent the 1-, 2-, and 3-$\sigma$ uncertainties of 
the LF estimated from the maximum-likelihood method. The blue 
and red dot-dashed curves show the LFs measured in the adjacent 
lower and higher redshift interval, to highlight the evolution with 
redshift of the LF. The dot-dashed cyan curve represents the local 
($z\sim0.1$) LFs in the rest-frame $^{0.1}r$ band (similar to our 
rest-frame $V$ band) from the SDSS \citep{montero09} and the GAMA 
survey \citep{loveday11}. The vertical arrows indicate the approximate 
90\% completenesses in rest-frame $V$-band magnitude of the catalogs 
used in our work. \label{fig2}}
\end{figure*}

\begin{figure}
\epsscale{1.}
\plotone{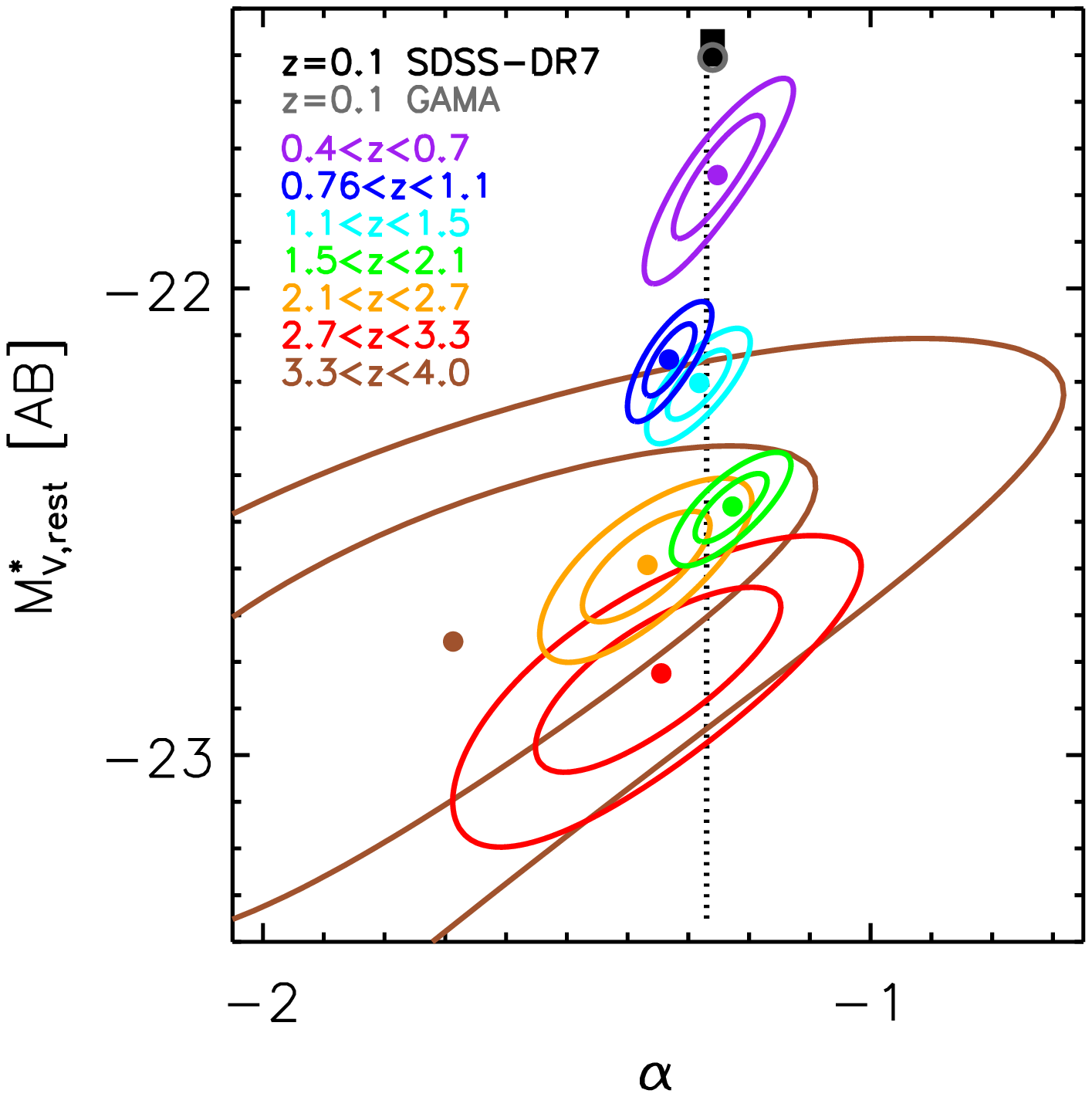}
\caption{($\alpha$-$M^{\star}_{\rm V}$) parameter space derived from the 
maximum likelihood analysis. Filled circles are the best-fit values of 
$\alpha$ and $M^{\star}_{\rm V}$ derived in our work, while the curves
represent their 1$\sigma$ and 2$\sigma$ contour levels. The black filled 
square represents the redshift $z\sim0.1$ value from the SDSS-DR7 
\citep{montero09}, whereas the empty gray circle represents the redshift 
 $z\sim0.1$ value from the GAMA survey \citep{loveday11}. The dotted line 
represents the weighted average of the plotted measurements 
($\alpha = −1.27\pm0.05$). \label{fig4bis}}
\end{figure}

The rest-frame $V$-band LFs of galaxies at $0.4\leq z<0.7$, $0.76\leq z<1.1$, 
$1.1\leq z<1.5$, $1.5\leq z<2.1$, $2.1\leq z<2.7$, $2.7\leq z<3.3$, and 
$3.3\leq z<4.0$ are shown in Figure~\ref{fig2}. Also plotted are the LFs in the 
rest-frame $^{0.1}r$ band (similar to our rest-frame $V$ band) at $z\sim0.1$ 
from the Sloan Digital Sky Survey Sixth Data Release (SDSS DR6; 
\citealt{montero09}) and the Galaxy and Mass Assembly (GAMA) survey 
\citep{loveday11}, to show the evolution from $z=0.55$ down to $z\sim0.1$. 
Note that the local LFs from \citet{montero09} and \citet{loveday11} are 
essentially identical. Figure~\ref{fig4bis} shows the evolution with redshift 
of the best-fit values and the 1~$\sigma$ and 2$\sigma$ confidence contour 
levels of the two Schechter function parameters $\alpha$ and $M^{\star}_{\rm V}$ 
derived with the maximum likelihood analysis.

The large surveyed area provided by the NMBS catalogs 
allows for the determination of the bright end of the rest-frame optical 
LFs of galaxies at $z>1$ with unprecedented accuracy, while the depth 
of the $H_{\rm 160}$-selected catalogs in UDF, HDFN, and ERS allowed us 
to constrain the faint-end slope out to $z\sim3.5$, representing a 
significant improvement with respect to previous works. Specifically, 
the NMBS samples the LF down to the characteristic magnitude at 
$z\sim3$, and down to $\sim 0.05~L^{\star}$ at $z\sim0.6$, while the ultra-deep 
$H_{\rm 160}$-selected catalogs allowed us to probe the faint end down to 
$\sim0.1~L^{\star}$ at $z\sim3.9$, and to $\sim0.003~L^{\star}$ at $z\sim1$. 
The approximate completenesses in rest-frame $V$-band magnitude of the 
catalogs used in our work are indicated in Figure~\ref{fig2} with arrows. 
Most importantly, the constructed NIR composite sample allowed us to measure, 
for the first time, the rest-frame optical LF of galaxies from $z=4.0$ using 
a single, self-consistent NIR composite sample that fully probes both the 
bright and the faint ends of the LF over the entire targeted redshift range 
$0.4 \leq z <4.0$. 

\begin{figure*}
\epsscale{0.9}
\plotone{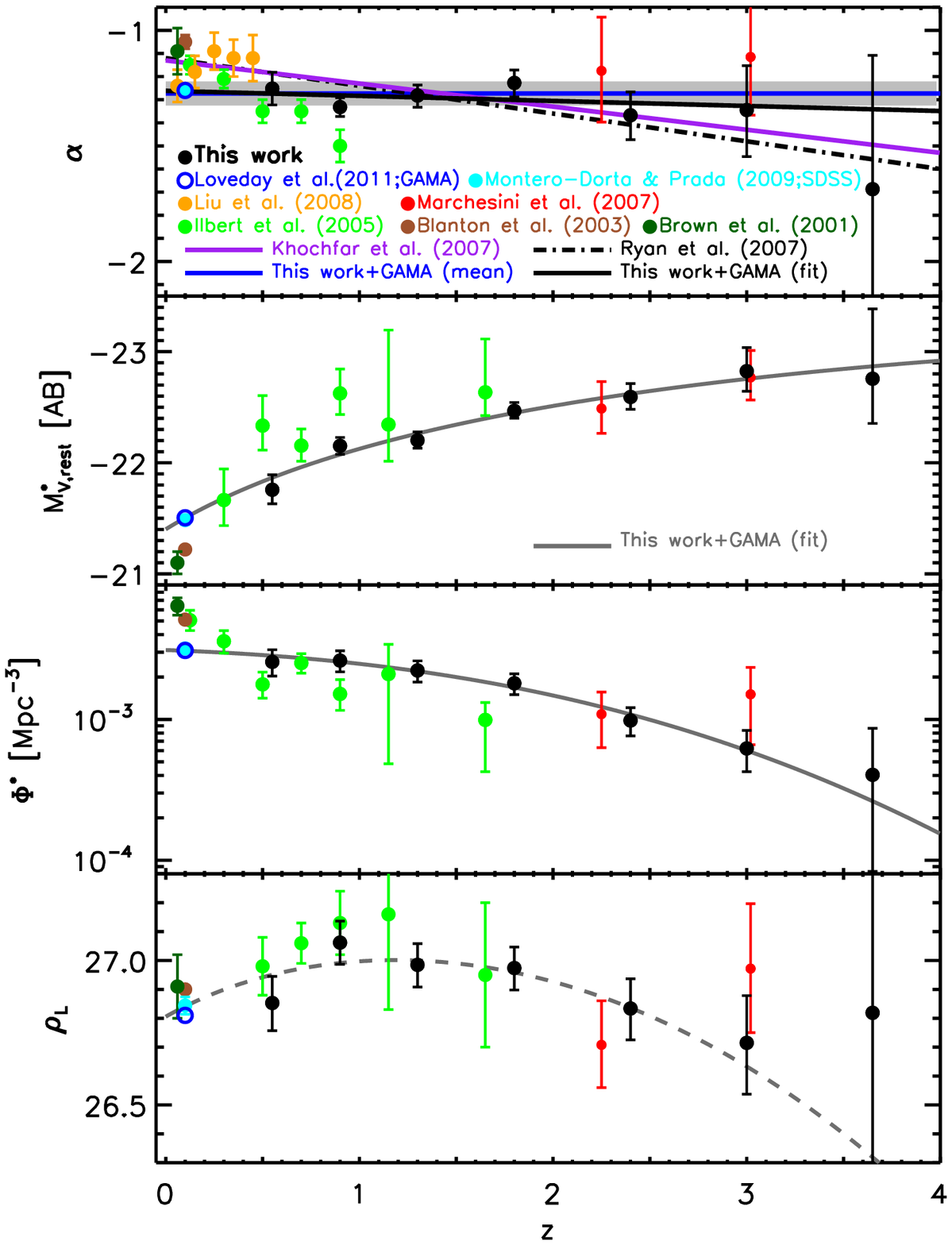}
\caption{Evolution of the \citet{schechter76} function parameters 
($\alpha$, $M^{\star}_{\rm V}$, and $\Phi^{\star}$, top three panels), and the 
luminosity density (bottom panel) with redshift. Filled black circles 
are the values derived in this work. Filled dark green, green, red, 
and orange circles are the values from \citet{brown01}, \citet{ilbert05}, 
\citet{marchesini07}, and \citet{liu08}, respectively. The open blue and 
filled cyan circles represent the rest-frame $^{\rm 0.1}r$-band (similar 
to our rest-frame $V$ band) results at $z\sim0.1$ from the GAMA survey 
\citep{loveday11} and the SDSS~DR6 \citep{montero09}, respectively. The 
filled brown circle represents the SDSS rest-frame $^{\rm 0.1}r$-band results 
at $z\sim0.1$ from \citet{blanton03}. The blue line in the top panel shows 
the weighted mean of the faint-end slopes from our work and 
\citet{loveday11}, with the gray shaded area representing the error of the 
mean. The purple line is the theoretical prediction from \citet{khochfar07}, 
whereas the dot-dashed black line is the best-fit model derived in 
\citet{ryan07}; the solid black line represents the best fit derived by 
fitting the values of $\alpha$ from \citet{loveday11} and our work with 
$\alpha(z)=a+bz$. The solid dark gray curves represent the best-fits of the 
evolution with redshift of $M^{\star}_{\rm V}$ and $\Phi^{\star}$ (using the 
parameterizations introduced in \citealt{stefanon11}; see \S~\ref{reslf}) 
using the values from \citet{loveday11} and our work. The dashed dark 
gray curve in the bottom panel is obtained adopting the best-fit curves of 
the evolution with redshift of $M^{\star}_{\rm V}$ and $\Phi^{\star}$ into 
Eq.~\ref{eq-rho}.\label{fig3}}
\end{figure*}

Table~\ref{tab-lf} lists the derived Schechter function parameters and the 
estimated luminosity densities in the targeted redshift intervals. 
Figure~\ref{fig3} shows the evolution with redshift of the Schechter function 
parameters $\alpha$, $M^{\star}_{\rm V}$, and $\Phi^{\star}$, as well as the 
evolution with redshift of the luminosity density. Figure~\ref{fig3} also 
shows the rest-frame $V$-band Schechter function parameters and luminosity 
densities derived in other works from the literature, i.e., the Century Survey 
\citep{brown01}, the SDSS \citep{blanton03}, the VIMOS-VLT Deep Survey (VVDS; 
\citealt{ilbert05}), the MUSYC-FIREWORKS-FIRES surveys \citep{marchesini07}, 
the Cosmic Evolution Survey (COSMOS; \citealt{liu08}), the SDSS-DR6 
\citep{montero09}, and the most recent GAMA survey \citep{loveday11}.

\begin{deluxetable*}{lclccccc}
\centering
\tablecaption{Best-fit Schechter Function Parameters and Luminosity Densities \label{tab-lf}}
\tablehead{
  \colhead{Redshift} & \colhead{$M^{\star}_{\rm V}$} & 
  \colhead{$\alpha$} & \colhead{$\Phi^{\star}$} &
  \colhead{$\log{\rho_{\rm L}}$} \\
  \colhead{}         & \colhead{(mag)}      & 
  \colhead{}         & \colhead{(10$^{-4}$~Mpc$^{-3}$~mag$^{-1}$)}   &
  \colhead{(erg~s$^{-1}$~Mpc$^{-3}$~Hz$^{-1}$)}  }
\startdata
$0.4 \leq z < 0.7$ & -21.76$^{+0.13}_{-0.14}$ & -1.25$\pm$0.07     & 25.61$^{+5.64}_{-5.33}$ & 26.85$^{+0.09}_{-0.10}$\\
$0.76\leq z < 1.1$ & -22.15$^{+0.08}_{-0.08}$ & -1.33$\pm$0.04     & 26.16$^{+4.53}_{-4.40}$ & 27.06$^{+0.07}_{-0.07}$\\
$1.1 \leq z < 1.5$ & -22.20$^{+0.07}_{-0.08}$ & -1.28$\pm$0.05     & 22.25$^{+3.79}_{-3.83}$ & 26.99$^{+0.07}_{-0.08}$\\
$1.5 \leq z < 2.1$ & -22.47$^{+0.07}_{-0.08}$ & -1.23$\pm$0.06     & 18.05$^{+3.02}_{-3.07}$ & 26.97$^{+0.07}_{-0.08}$\\
$2.1 \leq z < 2.7$ & -22.59$^{+0.11}_{-0.12}$ & -1.37$\pm$0.11     &  9.82$^{+2.28}_{-2.19}$ & 26.83$^{+0.10}_{-0.11}$\\
$2.7 \leq z < 3.3$ & -22.83$^{+0.18}_{-0.21}$ & -1.35$\pm$0.20     &  6.22$^{+2.12}_{-1.97}$ & 26.72$^{+0.16}_{-0.18}$\\
$3.3 \leq z < 4.0$ & -22.76$^{+0.40}_{-0.63}$ & -1.69$^{+0.58}_{-0.66}$ &  4.04$^{+4.60}_{-3.21}$ & 26.8$^{+1.5}_{-1.2}$\\
\enddata
\tablecomments{The quoted errors correspond to the 1~$\sigma$ 
errors estimated from the maximum likelihood analysis, including 
the contribution from cosmic variance and from photometric redshift 
uncertainties.}
\end{deluxetable*}

The faint-end slope $\alpha$ appears to evolve little with redshift. The 
weighted average of our measurements and the $z\sim0.1$ value from the GAMA 
survey (identical to the SDSS DR6) is $\alpha=-1.27\pm0.05$, shown in 
Figure~\ref{fig3} by the solid blue line and the shaded gray region. 
Adopting the simple linear fit presented in \citet{ryan07} ($\alpha(z)=a+bz$), 
we find $a=-1.26 \pm 0.02$ and $b=-0.02 \pm 0.03$ (plotted as a black solid 
line in Figure~\ref{fig3}) using the measurements from our work and from 
\citet{loveday11} in the fitting analysis. This is consistent with a constant 
faint-end slope in the rest-frame $V$ band over the redshift range $0<z<4$.
\footnote{If only the points from our work are used in the fitting analysis, 
we obtain $a=-1.30 \pm 0.06$ and $b=0.00 \pm 0.05$, with a resulting weighted 
mean $\alpha=-1.29\pm0.05$, quantitatively consistent with the analysis 
including the $z\sim0.1$ value.}

We have also modeled the evolution with redshift of the characteristic 
magnitude $M^{\star}_{\rm V}$ and the normalization $\Phi^{\star}$ using the 
parameterization presented in \citet{stefanon11}. Specifically, the following 
parameterization was adopted to model the observed evolution of 
$M^{\star}_{\rm V}$:

\begin{equation}
M^{\star}_{\rm V}(z)=\mu \left[ (1+z)/(1+z^{\star}) \right]^\eta \exp{\left[-(1+z)/(1+z^{\star})\right]},
\label{eq:sch_ms}
\end{equation}
with $\mu$, $z^{\star}$, and $\eta$ free parameters to be determined. By 
performing a least-square fit to our measurements and from \citet{loveday11}, 
we obtain the following values: $\mu=-28.8\pm 0.4$ mag, $z^{\star}=191\pm 338$, 
and $\eta=(55.5\pm 17.2) \times 10^{-3}$. The resulting curve is plotted as a 
solid dark gray curve in Figure~\ref{fig3}. The characteristic magnitude is 
seen to brighten with redshift, from $M^{\star}_{\rm V}=-21.5$~mag at $z\sim 0.1$ 
to $M^{\star}_{\rm V}=-22.8$~mag at $z\sim3.7$.

The following parameterization was adopted to model the observed evolution 
of $\Phi^{\star}$:

\begin{equation}
\Phi^{\star}(z)= \theta \exp{\left[\gamma/(1+z)^\beta\right]},
\label{eq:fit_phi}
\end{equation}
where $\theta$, $\gamma$, and $\beta$ are the free parameters. The best-fit 
values are $\theta= (3.2 \pm 0.4) \times 10^{-3}$ mag$^{-1}$ Mpc$^{-3}$, 
$\gamma=(-4.3 \pm 5.0) \times 10^{-2}$, and $\beta=-2.65 \pm 0.88$. The 
best-fit of Equation \ref{eq:fit_phi} is plotted as a solid dark gray curve in 
Figure~\ref{fig3}. The normalization is seen to increase with cosmic time, 
by a factor of $\sim6^{\rm +32}_{\rm -4}$ from $z\sim3.7$ to $z\sim0.6$, and 
by a factor of $\sim8^{\rm +33}_{\rm -4}$ to $z\sim0.1$, with 50\% of such 
increase in the 2~Gyr from $z=4$ to $z=1.8$, and the remaining increase 
in the following 9~Gyr from $z=1.8$ to $z=0.1$.

Using the expressions of Eq.~\ref{eq:fit_phi} and Eq.~\ref{eq:sch_ms} in 
Eq.~\ref{eq-rho}, we can derive a functional representation of the luminosity 
density. The dashed dark gray curve in the bottom panel of Figure~\ref{fig3} 
represents the luminosity density obtained with this method and adopting the 
best-fit values of the parameters previously recovered. The agreement with 
the points is good over the entire redshift range. We stress that no fitting 
was performed using the luminosity density estimates. The luminosity 
density appears to peak at $z\approx 1-1.5$, increasing by a factor of 
$\sim4$ from $z\sim3.7$ to $z\sim1-1.5$, and subsequently decreasing by a 
factor of $\sim1.5$ to $z\sim0.1$.


\section{Summary and Conclusions} \label{conc}

In this paper, we have measured the rest-frame $V$-band LFs of galaxies 
at $0.4 \leq z < 4.0$ from a composite NIR-selected sample constructed 
with the deep NMBS, the FIREWORKS, the very deep FIRES, and the ultra-deep 
HST optical and NIR data over the HUDF, HDFN, and GOODS-CDFS fields, all 
having high-quality optical-to-MIR data. This sample is unique as it 
combines, in a self-consistent way, the advantages of deep, pencil beam 
surveys with those of shallow, wider surveys, allowing us to (1) minimize 
the uncertainties due to cosmic variance, and empirically quantify its 
contribution to the total error budget; and (2) simultaneously constrain 
the bright and faint ends of the rest-frame optical LFs with unprecedented 
accuracy over the entire targeted redshift range.

The main results of our derivation of the rest-frame $V$-band LF of 
galaxies from $z=4.0$ are (1) the faint-end slope remains fairly flat 
and does not seem to evolve from $z=4$ to $z=0$, consistent with 
$\alpha=-1.27\pm0.05$; (2) the characteristic magnitude has dimmed 
by 1~mag from $M^{\star}_{\rm V}=-22.8$ at $z\sim3.7$ to $M^{\star}_{\rm V}=-21.8$ 
at $z\sim0.6$, and by $\sim1.3$~mag to $z=0.1$; (3) the normalization 
$\Phi^{\star}$ has increased by a factor of $\sim6$ from $z\sim3.7$ to 
$z\sim0.6$, and by a factor of $\sim8$ to $z=0.1$; and (4) the luminosity 
density peaks at $z\approx 1-1.5$, increasing by a factor of $\sim4$ from 
$z=4.0$ to $z\approx 1-1.5$, and subsequently decreasing by a factor of 
$\sim1.5$ by $z=0.1$.

In contrast with the findings of \citet{ryan07}, we find weak, if any, 
evidence for a steepening of the faint-end slope with redshift out to $z=4$. 
The top panel of Figure~\ref{fig3} shows the best-fit model derived in 
\citet{ryan07} (dot-dashed line). Using the simple linear fit $\alpha(z)=a+bz$, 
\citet{ryan07} found $b \approx -0.12$, whereas we find $b=-0.02 \pm 0.03$, 
consistent with no evolution with redshift of the rest-frame $V$-band 
faint-end slope over the redshift range $0.4 \leq z < 4.0$. We note that the 
evolution of the faint-end slope with redshift presented in \citet{ryan07} 
was found by combining measurements of the faint-end slopes in the rest-frame 
$B$ and UV bands. When only the rest-frame $B$-band values are considered, 
the evolution of the faint-end slope with redshift becomes marginal. While 
the rest-frame $B$ and $V$ bands are close enough that significant 
differences should not be expected, we cannot exclude systematic differences 
in the (evolution) of the faint-end slopes in different rest-frame optical 
bands. It should be finally noted that a steepening in the faint-end slope 
at $z \gtrsim 2.5$ might still be present, due to the large uncertainties 
on the faint-end slope at $z>3$.

The hierarchical formation scenario states that many dwarf galaxies at high 
redshifts will undergo merging over time. This implies a steeper faint-end 
slope at higher redshifts. Using semi-analytical modeling of galaxy formation, 
\citet{khochfar07} investigated the evolution with redshift of the rest-frame 
$B$-band faint-end slope of the LF, finding $b \approx -0.1$, in good 
agreement with the results from \citet{ryan07}. The predicted evolution of 
$\alpha$ from \citet{khochfar07} is also plotted in Figure~\ref{fig3} (purple 
line). Our measurements are consistent with no evolution of the faint-end 
slope over the last 12~Gyr of cosmic history, in apparent contrast with what 
is expected in the current paradigm of structure formation. Supernova feedback 
is thought to be mainly responsible for a flatter faint-end slope compared to 
the dark matter halo mass slope $\alpha_{\rm DM}$ at any given redshift 
\citep{dekel86}, with $\alpha_{\rm DM}$ flattening with cosmic time, as shown 
by $\Lambda CDM$ numerical simulations (e.g., \citealt{reed03}). A constant 
$\alpha$ from $z=4$ would then imply either no evolution in $\alpha_{\rm DM}$ 
in the last 12~Gyr, in contrast with $\Lambda CDM$ simulations, or a 
redshift-dependent efficiency of stellar feedback, with supernova feedback 
being more efficient at higher redshifts when $\alpha_{\rm DM}$ is also 
steeper. Alternatively, the non-evolving rest-frame $V$-band faint-end slope 
out to $z=4$ could be explained while maintaining a constant stellar feedback 
efficiency, if we assumed that the galaxies at the faint end are hosted by 
different dark matter halos at different redshifts. In other words, the 
observed faint galaxies at low redshifts are hosted by more massive dark 
matter halos compared to similarly faint galaxies at high redshifts. 
Additionally, systematic effects could be introduced by the sampled 
environments of faint galaxies, if these environments vary systematically 
at different redshifts. In fact, as shown by \citet{khochfar07}, 
$\alpha$ is steeper and evolves more strongly in field environments than in 
cluster environments. Consequently, if the faint-end of the LF at high 
redshifts is observationally dominated by present-day cluster members, the 
faint-end slope of the overall galaxy population at high redshifts would 
then be flatter than that for the field LF alone at the same redshift, 
potentially resulting in the observed little or no evolution with redshift 
of $\alpha$.

Whereas we find a fairly flat and constant faint-end slope from $z=4$ in 
the rest-frame $V$ band, significantly steeper slopes are consistently found 
in the rest-frame far-UV, both at $z>4$ ($\alpha \lesssim -1.7$; e.g., 
\citealt{bouwens07}; \citealt{ouchi09}; \citealt{oesch10a}; 
\citealt{bouwens11}; \citealt{lee11}) and at $z<4$ ($\alpha \sim -1.6$; 
\citealt{arnouts05}; \citealt{reddy09}; \citealt{hathi10}; 
\citealt{oesch10b}). Our NIR composite sample will allow us to study the 
evolution of the LFs from $z=4$ at different rest-frame wavelengths, from 
the far-UV to the NIR, and to investigate the wavelength-dependence of the 
faint-end slope and its potentially different varying evolution with redshift 
in different rest-frame wavebands.

We stress that we have measured the {\it observed} rest-frame $V$-band LFs 
over the targeted redshift range $0.4 \leq z < 4.0$. No attempt was made to 
infer the intrinsic, dust-corrected, rest-frame $V$-band LF of galaxies, 
which is beyond the scope of our work. We note however that, similar to what 
has been found in the local universe (e.g., \citealt{driver07}), internal dust 
extinction is likely to play a significant role in shaping the observed 
LF, especially in the early universe where larger amount of dust extinction 
is typically found, even in galaxies around the characteristic mass (e.g., 
\citealt{marchesini10}; \citealt{brammer11}).

The ongoing {\it Hubble Space Telescope} CANDELS (\citealt{grogin11}; 
\citealt{koekemoer11}) and 3D-HST (\citealt{vandokkum11}; \citealt{brammer12}) 
surveys, once completed, will result in significantly improved constraints at 
the faint end, providing 1) very deep data over an area larger by a factor 
$> 10$ with respect to the area surveyed by the ultra deep HST data used in 
this work, and 2) spectroscopic redshifts for thousands of faint galaxies, 
significantly reducing the uncertainties due to photometric redshift errors. 


\acknowledgments
We thank the anonymous referee for helpful comments and suggestions 
that significantly improved the manuscript. We thank Lars Hernquist 
for helpful discussions. This study makes use of data from the NEWFIRM 
Medium-Band Survey, a multi-wavelength survey conducted with the NEWFIRM 
instrument at the KPNO, supported in part by the NSF and NASA. The authors 
acknowledge support from programs HST-AR-11764.04 and HST-AR-12141.01, 
provided by NASA through a grant from the Space Telescope Science Institute, 
which is operated by the Association of Universities for Research in 
Astronomy, Incorporated, under NASA contract NAS5-26555.
Based on observations made with ESO Telescopes at the La Silla or 
Paranal Observatories under programme ID(s): 66.A-0270, 67.A-0418, 
074.A-0709, 164.O-0560, 170.A-0788, 171.A-3045, and 275.A-5060.


\end{document}